\documentclass[epj]{svjour}
\usepackage{xurl} 
\usepackage{graphicx} 
\usepackage{amsmath}
\usepackage{xcolor}   
\usepackage{listings} 
\usepackage{booktabs} 
\usepackage[hidelinks]{hyperref} 

\begin{document}
\title{GPU acceleration of optical photon propagation in low photon yield applications: Opticks for the Electron Ion Collider}

\author{Gabor Galgoczi\inst{1}\thanks{Corresponding author: \texttt{ggalgoczi1@bnl.gov}}
\and Kolja Kauder\inst{1}
\and Maxim Potekhin\inst{1}
\and Sakib Rahman\inst{1}
\and Dmitri Smirnov\inst{1}
\and Torre Wenaus\inst{1}
}                     
\offprints{\texttt{ggalgoczi1@bnl.gov}}          
\institute{Brookhaven National Laboratory, Upton, NY 11973, USA}
\date{December 2025}
%
\abstract{
The bulk of time spent in the simulation of Cherenkov and other scintillation detectors is spent on optical-photon transport, i.e. \emph{ray tracing}, a task that GPUs are uniquely qualified to perform. We present EIC-Opticks, a fork of Opticks, which uses event aggregation to drastically accelerate photon transport simulation for low-to-moderate photon yield experiments. During the full Geant4 Monte Carlo simulation of a given detector, optical photon simulation is performed on GPU(s) using the NVIDIA OptiX framework.
We validate this approach using the ePIC pfRICH detector. We find GPU and CPU simulations in excellent agreement. For $5\times 10^4$ electrons with a momentum of $p=5~\mathrm{MeV}/c$ in the test case of the pfRICH detector, EIC-Opticks shows an order-of-magnitude speedup over multi-threaded Geant4, and a factor of up to 161$\pm$3 over single-threaded execution. In the case of low-to-moderate applications event aggregation reduces the per-photon simulation time from $\sim60\,\mu\mathrm{s}$ for single events to $\sim20\,\mathrm{ns}$ with batching, a factor of $\sim3000$. 
In order to make EIC-Opticks easily installable, we authored a Spack package that makes it possible to install it with a single command.
Additionally, a Docker container is provided for users with EIC-Opticks installed. EIC-Opticks provides guardrails for common pitfalls (e.g. nested volume conversion, ray tracing setting optimization).}
%
%
\maketitle

\section{Introduction}

Optical photon simulation is a critical, yet extremely time-consuming part of scintillator and Cherenkov photon detector simulations in both high-energy (HEP) and nuclear physics (NP). In these fields, the most widely used simulation tool is Geant4~\cite{geant4,geant4-2016}, which executes optical processes on the CPU. To give an example for the JUNO detector, 99\% of simulation time was spent solely on the simulation of optical photons~\cite{opticks-chep2018}. This limits the size of simulated samples, slows design iterations, and complicates end-to-end studies that must be repeated across many geometries. Furthermore, it limits the size of output data that can be used for Machine Learning applications.

At the same time, specialized Graphics Processing Units (GPUs) have gained massive popularity in entertainment and professional applications. Billions of dollars have been spent since the start of the millennium with no sign of slowing down. The key technologies are massive vectorization and since ca. 2018 dedicated ray tracing hardware. However, algorithms natively available for GPUs do not support processes such as Rayleigh scattering. The Opticks software~\cite{opticks-chep2018,opticks-chep2023} was built NVIDIA's ray-tracing engine OptiX~\cite{optix-2010}. Opticks also incorporates pertinent physics like Mie scattering, Rayleigh scattering, etc.
Monte Carlo methods are used to introduce the randomness needed in HEP and NP simulations. Opticks was originally developed for the JUNO experiment, where single events generate hundreds of millions of optical photons. These photons occupy the entire RAM of the GPU. Thus, overhead is minimal. However, most high-energy and nuclear physics experiments involving optical photons yield only a few hundred photons per individual event. Therefore, na\"ive per-event offloading can be slower than CPU-only simulation because the fixed overhead per offload dominates the simulation time.

To address this issue, we introduce EIC-Opticks\footnote{https://github.com/BNLNPPS/eic-opticks}, a fork and extension of Opticks.
 Although developed mainly for Electron Ion Collider (EIC) \cite{eicyellow-2022} detector studies, it lends itself readily to many low-to moderate photon-yield applications, including medical imaging simulations. EIC-Opticks adds an event aggregation mode that batches thousands of Geant4 events into a single GPU call. This design retains physics fidelity while removing per-event overheads that otherwise erase GPU advantages at low photon counts. 

The framework supports the standard optical processes required for Cherenkov detectors (e.g. wavelength-dependent refractive indices and absorption, Rayleigh scattering, and configurable surface reflectance/roughness models), and it integrates with common HEP/NP workflows.
For usability and reproducibility, we provide containerized distributions\footnote{\url{https://github.com/BNLNPPS/esi-shell}} and a Spack~\cite{spack-sc15} package, enabling turnkey deployment on hosts with modern NVIDIA GPUs.

We validated EIC-Opticks using the \emph{pfRICH} detector \cite{eicyellow-2022} as a representative EIC subsystem. In a study with $5\times 10^4$ electrons at $p=5~\mathrm{MeV}/c$, we compare observables between \textsc{Geant4} and EIC-Opticks and find agreement. In the same configuration, batching reduces runtime dramatically: relative to single-threaded \textsc{Geant4} we observe speedups exceeding two orders of magnitude in the batched case, while outperforming multi-threaded \textsc{Geant4} by a factor of ten for typical pfRICH workloads. These gains make previously prohibitive optical studies practical at scale and open the door to systematic scans over geometry, surface treatments, and operating conditions, not to mention the order of magnitude more training data for Machine Learning applications.

The paper is structured as follows: Section~\ref{sec:mod} describes EIC-Opticks and the batching strategy. Section~\ref{sec:pfrich} summarizes the pfRICH setup used in this work. Section~\ref{sec:validation} presents physics validation against \textsc{Geant4}. Section~\ref{sec:performance} reports performance benchmarks and configuration studies. Section~\ref{sec:leakage} discusses precision limits and mitigation strategies. We conclude with a summary.

\section{EIC-Opticks}
\label{sec:mod}

Opticks is a GPU-accelerated framework that couples to \textsc{Geant4} and offloads optical-photon transport to NVIDIA’s \textsc{OptiX} ray-tracing runtime while leaving all non-optical physics on the CPU. Building on this model, EIC-Opticks retains the hybrid execution workflow and introduces an event–aggregation mode that batches thousands of low yield \textsc{Geant4} events into a single GPU simulation. This removes per-event launch and transfer overheads without changing the underlying physics and makes configurations such as pfRICH efficient even when individual events produce only $\mathcal{O}(10^2\text{–}10^4)$ photons per event.

To protect users from crashes due to failed geometry conversions, EIC-Opticks analyzes constructive solid geometry (CSG) upon conversion from the input GDML~\cite{gdml-2006} files and emits warnings when it detects deeply nested boolean operations (unions, subtractions, intersections). Since CSG is represented as full binary trees in the accelerator architecture (GPU representation of the geometry), such nested boolean operations can yield a binary tree size of millions of leafs. In such cases, we suggest modifications to the volume in question.

For usability and reproducibility, we migrated the most important runtime controls from shell environment variables into a versioned configuration file. Parameters include the maximum number of boundary reflections, and the ``propagation epsilon''~\cite{pbrt-managing-error,nvidia-dxr-self-intersection}, probably the most important user-defined setting in the NVIDIA OptiX framework (see Section~\ref{sec:leakage} for further details).

 EIC-Opticks is distributed as a container image for zero-install execution and provides a \textsc{Spack} package for managed builds, ensuring consistent deployment across developer workstations and computer clusters. As an additional option EIC-Opticks can be installed with the \textsc{Spack} commands.

We also provide geometry specific tools to optimize crucial settings. 
First, the maximum number of reflections controls when a photon is discarded. 
If the setting is too large, a small portion of photons that do not contribute significantly to the hit numbers use up excessive GPU time. If it is too small, a significant number of relevant hits may be missing. To optimize this number, simulations can be automatically run with several settings until the number of hits converges. 
This analysis is needed for each geometry, since the expected mean path and detector topology varies case by case. In Section~\ref{sec:pfrich} we illustrate this trade-off specifically for the pfRICH geometry. 

Secondly, the ray-tracing ``epsilon'' is a parameter of the underlying NVIDIA OptiX simulation. It defines by how much each step is shifted on purpose to avoid self intersection. If epsilon is too large, then photons may teleport through a boundary surface. If it is too small, a photon may get very close to a boundary and again teleport through, this time due to rounding errors. We will demonstrate details of this trade-off again in the context of the pfRICH in Subsection~\ref{sec:leakage}.

\section{pfRICH detector at the Electron Ion Collider}
\label{sec:pfrich}

The proximity-focusing Ring-Imaging Cherenkov Detector (pfRICH) is currently being constructed as part of the Electron-Proton/Ion Collider (ePIC) \cite{eicyellow-2022} detector system slated to start operations in the mid-2030s
Designed to provide pion-kaon separation up to 7 GeV/c \cite{chatterjee_epic_pid_2025}, the pfRICH detector subsystem consists of a thin radiator layer, a 40 cm proximity gap, a photosensor plane, and two mirrors to increase acceptance. The optical design prioritizes compactness and low material budget while maintaining the high pion-kaon separation, which makes it ideal to stress-test the EIC-Opticks software.

\begin{figure}[h!]
    \centering
    \includegraphics[width=0.5\textwidth]{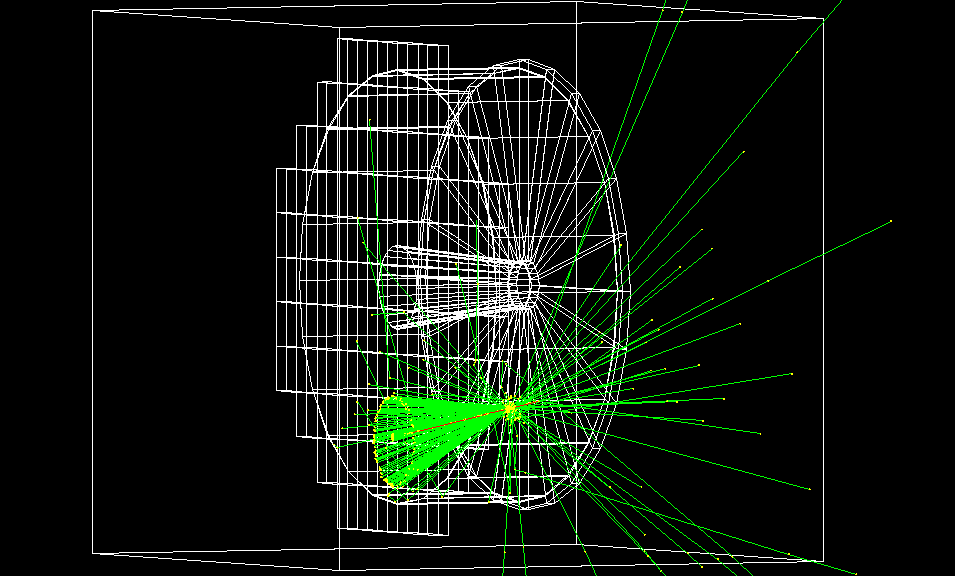}
    \caption{The Geant4 simulation of a single 5~MeV electron (red track) entering the pfRICH detector, creating Cherenkov photons (green tracks) mainly in the aerogel tiles. }
    \label{fig:pfrich_pic}
\end{figure}

The collaboration’s original GDML file defining the pfRICH detector consists of 126 volumes. For simulation studies in this work, we use a minimally simplified geometry that preserves all optically relevant elements: aerogel radiators (with wavelength-dependent refractive index and Rayleigh scattering), inner/outer mirror assemblies with defined optical surfaces, the photosensor window and detection planes. Volumes that are irrelevant to photon transport or that rely on Geant4 solids currently unsupported (trapezoids) in our analytic GPU geometry description are substituted with optically similar constructs that match surface normals and refractive/reflective interfaces. In Fig.~\ref{fig:pfrich_pic} a 5~MeV electron is shown with additional photons created by the Cherenkov process.

\section{Validation}
\label{sec:validation}

In order to validate EIC-Opticks we have used the exact same geometry both in a standalone Geant4 simulation (version 11.2.1) and in a simulation where the optical photons were offloaded to EIC-Opticks to be simulated on GPU. We used the GDML file of the pfRICH geometry, with the only modification being the approximated trapezoid volumes mentioned above. Note that for the purposes of this comparison, the standalone Geant4 simulation uses the same geometry as EIC-Opticks. The number of maximum reflections was set to 24 in EIC-Opticks; this choice is explained below in Section~\ref{sec:performance}.

50,000 electrons with a momentum of 5 MeV/c were simulated in individual events. In the case where all particles were simulated on Geant4 (CPU) 8,693,457 $\pm$ 2948  hits were recorded. When optical photons were offloaded onto the GPU to be simulated by EIC-Opticks 8,693,758 $\pm$ 2949 hits were recorded.

\begin{figure}[h]
    \centering
    \includegraphics[width=0.5\textwidth]{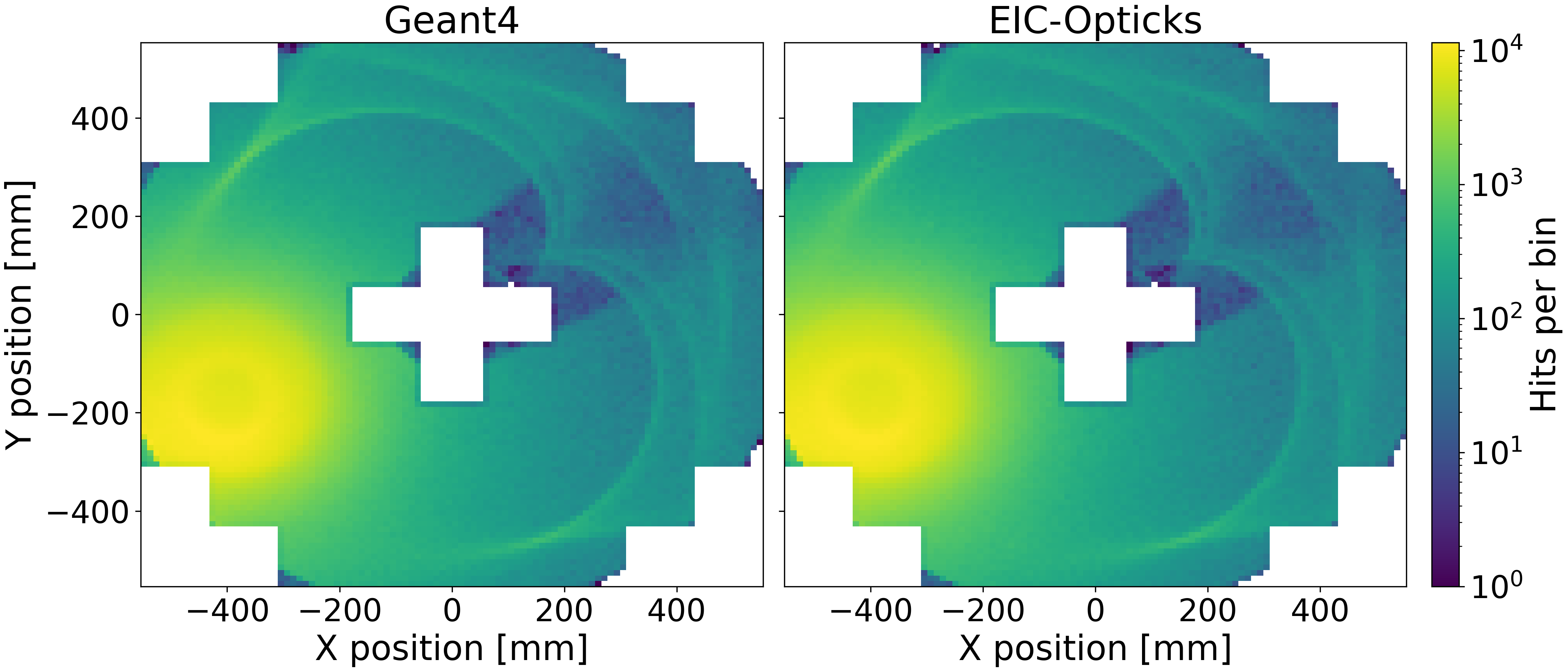}
    \caption{50,000 electrons with momentum 5 MeV/c shot in as a pencil beam. Hit maps simulated by Geant4 (CPU) validation on the left and EIC-Opticks (GPU) on the right. Event batching introduced in EIC-Opticks was utilized. The identical (within statistical errors) hit maps prove that EIC-Opticks yields the same results as Geant4.}
    \label{fig:gdmlpf}
\end{figure}

In Fig. \ref{fig:gdmlpf} the hit histograms of generated photons are shown, the left panel was generated by regular Geant4 on a CPU, and the right one with EIC-Opticks. One can see that the histograms are identical, and the Cherenkov ring and caustic curves generated by the reflection of the optical photons on the outer mirror are visible in both.


\section{Performance}
\label{sec:performance}

\begin{figure}[h!]
    \centering
    \includegraphics[width=0.5\textwidth]{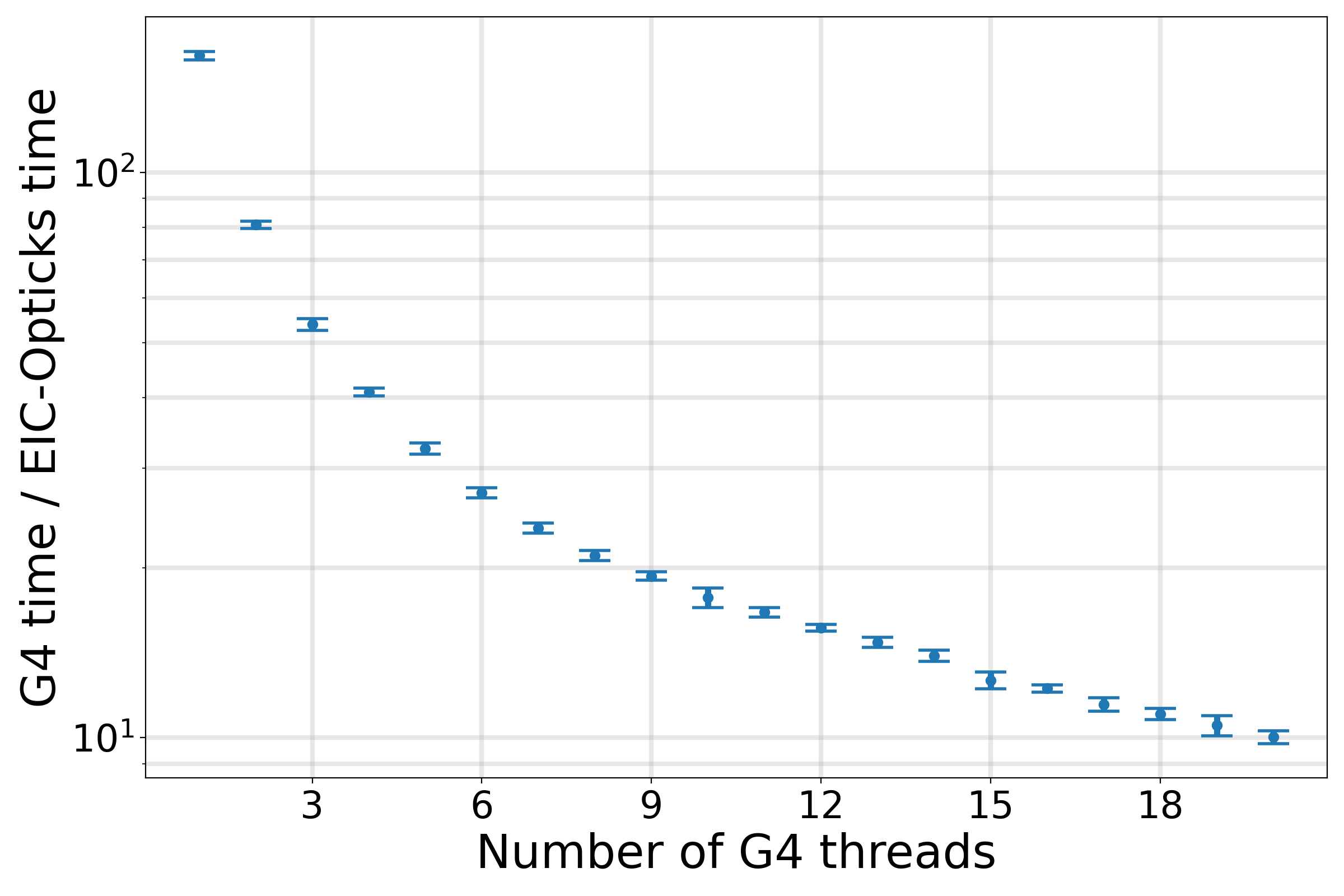}  
    \caption{Speed increase of the GPU simulation compared to propagating the photons with multi-threaded Geant4 (CPU) for different number of Geant4 threads.}
    \label{fig:eicopticksvsg4}
\end{figure}

In order to compare the simulation of EIC-Opticks to Geant4 optical photon propagation, 50,000 electrons, each with a momentum of 5 MeV/c were simulated yielding 44 million optical photons. The time required to simulate the photons in Geant4 was measured as the time difference  between the full simulation run with and without optical photon propagation. In the case of EIC-Opticks, we included all the overhead: copying the would-be simulated data to GPU, ray-tracing, and copying the results back to memory.  In Fig.~\ref{fig:eicopticksvsg4} the speedup using EIC-Opticks is plotted with respect to a multi-threaded Geant4 optical photon propagation. The GPU simulation is 161$\pm$3 times faster compared to a single thread of Geant4 and 10.0$\pm$0.3 times faster than a multi-threaded Geant4 simulation with 20 threads. The GPU utilized was an NVIDIA GeForce RTX 4090, and the CPU was an Intel Xeon w7-3445. 

\begin{figure}[h!]
    \centering
    \includegraphics[width=0.45\textwidth]{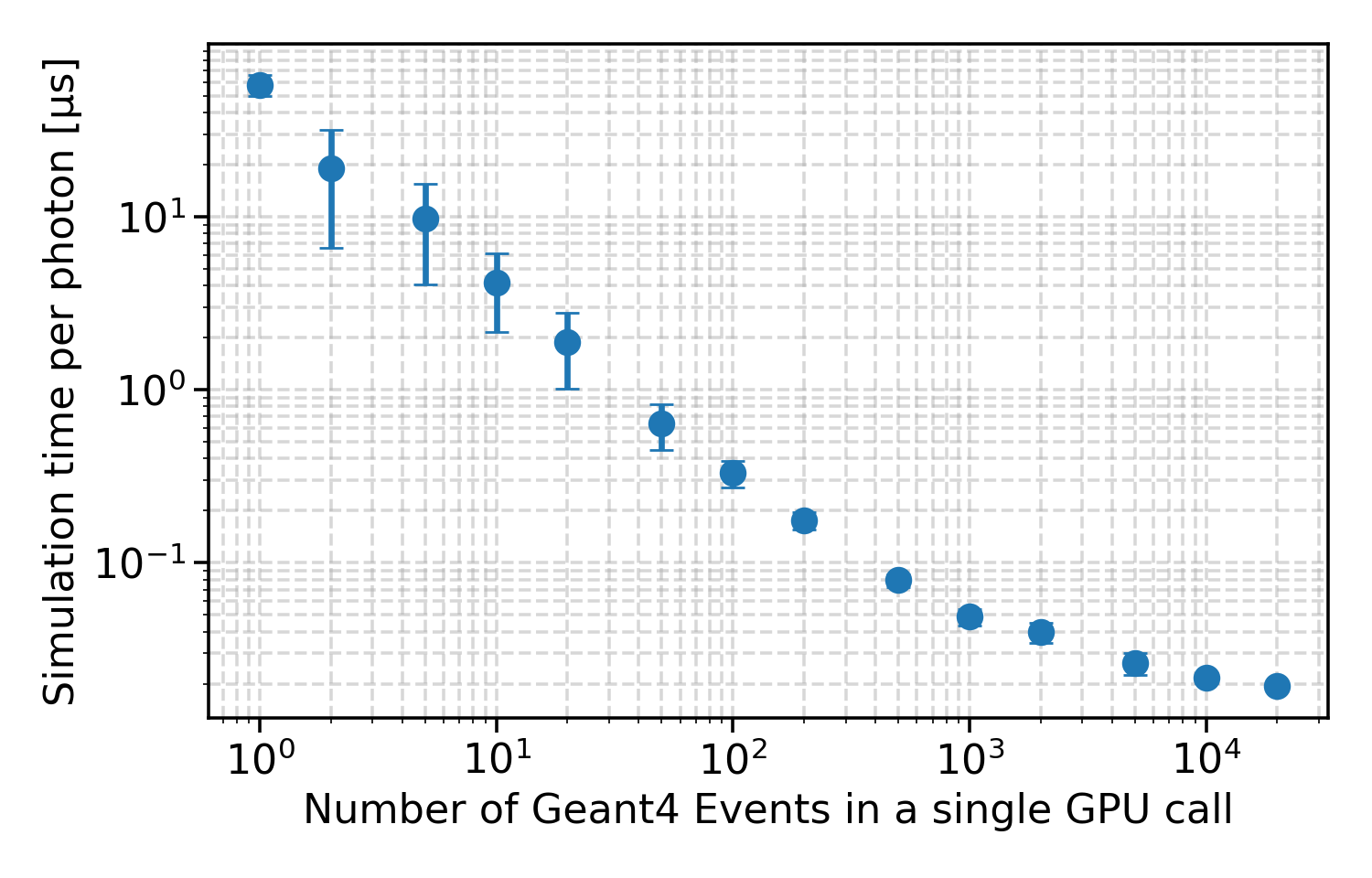}%
    \caption{Average time spent on the simulation of a single photon with EIC-Opticks
    for different event batch sizes in a single GPU call.}
    \label{fig:batching1}
\end{figure}

\begin{figure}[h!]
    \centering
    \includegraphics[width=0.45\textwidth]{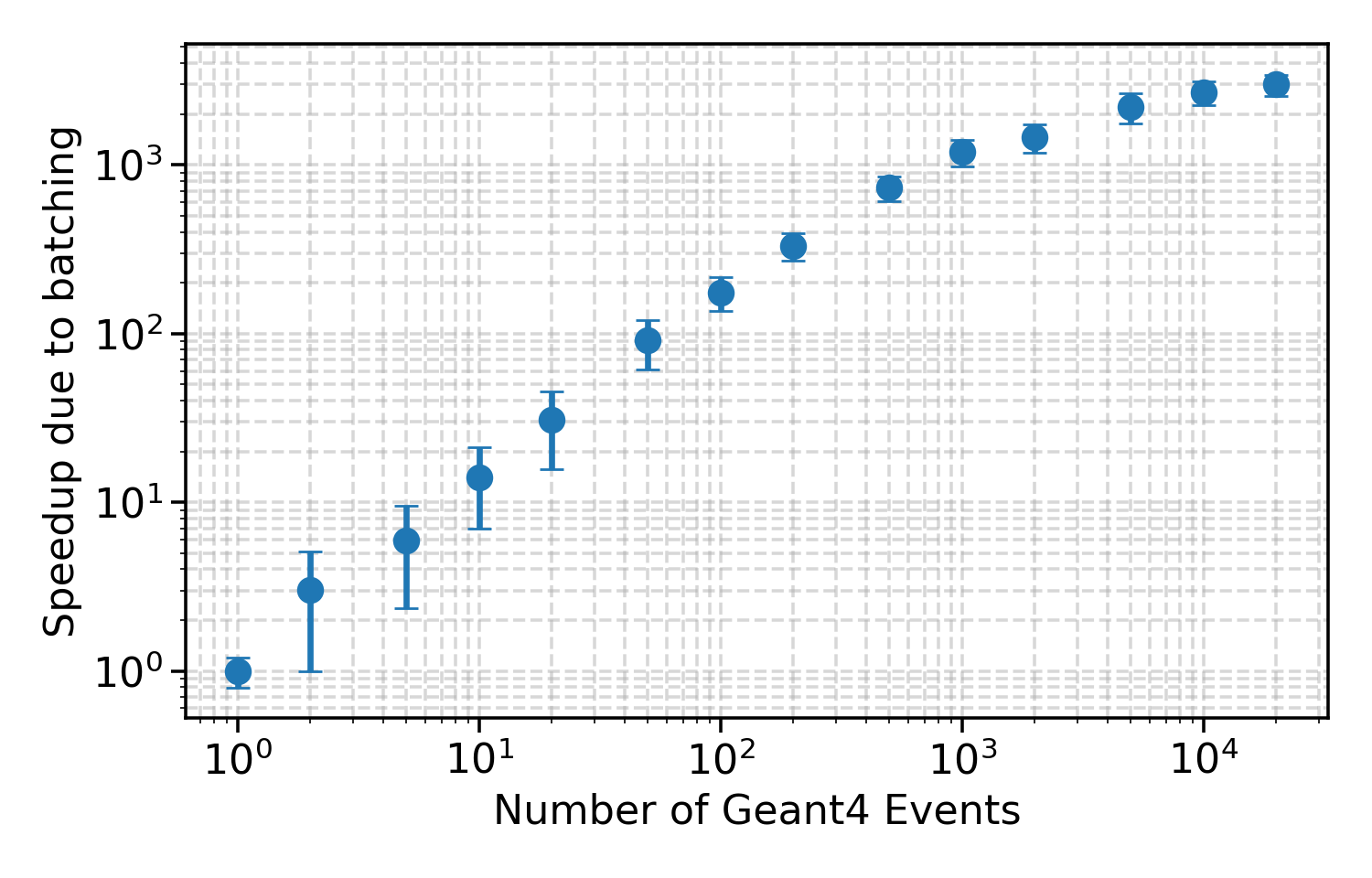}%
    \caption{Speed increase of the GPU simulation due to batching compared to no batching.
    For the pfRICH detector, simulating 10{,}000 events in a single GPU call yields a speedup
    of three orders of magnitude.}
    \label{fig:batching2}
\end{figure}

The pfRICH's photon occupancy is relatively small, meaning that overhead plays a significant role on a per-event basis, and furthermore that the GPU is not used to its full potential. We can achieve an additional dramatic speed-up by batching multiple Geant4 events into a single GPU launch.
In the left panel of Fig.~\ref{fig:batching1}, the average time required to propagate a single photon is plotted for different numbers of Geant4 events batched together on the GPU.
When a single event with $\sim$100 photons is simulated, it takes $\sim$60$~\mu$s on average to propagate one photon.
Meanwhile, a batch of 10,000 events reduces this number  to $\sim$20~ns. On the right of Fig.~\ref{fig:batching2}, the speedup is shown for different numbers of events batched together compared to no batching. With the GPU's memory thus fully utilized, a speedup of $\sim$3000 is reached.

\begin{figure}[h]
    \centering
\includegraphics[width=0.5\textwidth]{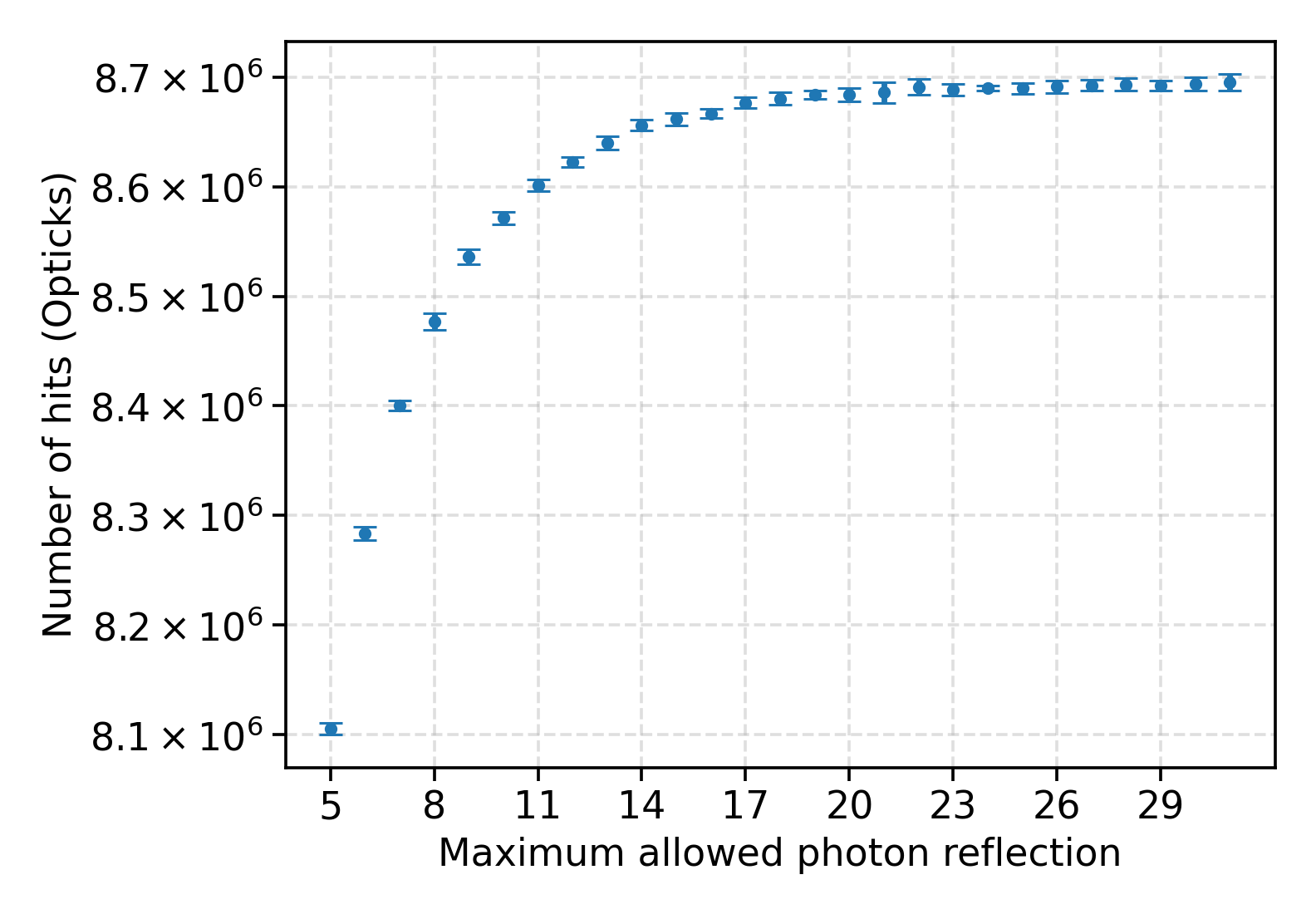}  
    \caption{Number of hits in the pfRICH geometry upon simulating 50,000 electrons 5 MeV/c with maximum reflection settings from 5 up to 32. One can see that the number of hits plateaus around $\approx$24, in which case the number agrees with the baseline CPU Geant4 simulation within a statistical error.}
    \label{fig:bounce}
\end{figure}

The maximum number of reflections on surfaces is one of the most important settings in EIC-Opticks, described in Section \ref{sec:leakage}. We simulated 50,000 electrons with momentum of 5 MeV/c with maximum reflection settings from 5 up to 32 in steps of 1. In Fig. \ref{fig:bounce} the number of hits can be seen for different reflection settings. The number of hits starts to plateau around $\sim$24 reflections.

In order to properly measure the run time and see whether there is a significant difference for specific settings, 20 simulations each were run for maximum reflection values of 8, 16, 24, and 32. Table~\ref{tab:bounces} shows the simulation time for each setting. The time variation for different settings is on the order of 10\%. However, for specific use cases, e.g. elongated scintillator bars where photons can be reflected hundreds of times, warp---GPU core group that shares instructions---divergence still has to be investigated.

\begin{table}[ht]
\centering
\caption{GPU simulation time required to simulate 50,000 electrons with a momentum of 5 MeV/c for multiple maximum reflection settings}
\label{tab:bounces}
\begin{tabular}{cc}
\toprule
\textbf{Maximum Reflections} & \textbf{Time [s]}  \\
\midrule
8  & 0.737 $\pm$ 0.008 \\
16 & 0.83 $\pm$ 0.01 \\
24 & 0.86 $\pm$ 0.01 \\
32 & 0.86 $\pm$ 0.01 \\
\bottomrule
\end{tabular}
\end{table}

\begin{table}[htbp]
  \centering
  \caption{Simulation time of optical photons created by 50 000 e$^{-}$ with momentum of 5 MeV/c in the pfRICH geometry. }
  \label{tab:performance}
  \begin{tabular}{l c}
    \toprule
    Method                & Time (s)  \\
    \midrule
    Geant4 single thread  & $\sim 120$ \\
    Geant4 20 threads     & $\sim 8$  \\
    Opticks              & $\sim 0.8$ \\
    \bottomrule
  \end{tabular}
\end{table}

\section{Photon leakage}
\label{sec:leakage} 

EIC-Opticks uses ray-tracing provided by the underlying NVIDIA OptiX framework. When the track of a photon crosses a boundary, na\"ively the next step should start at the boundary. However, OptiX utilizes single precision numbers, and when calculating the origin of the next step, there is a chance that it will lie outside the volume. Thus, instead of being reflected on the surface in the second step, the ray will intersect the boundary from the outer side and thus escape from the volume, in other words, ``leak'' out. This happens due to the reflection process that still changes the direction of the ray.

\begin{figure}[h!]
    \centering
    \includegraphics[width=0.5\textwidth]{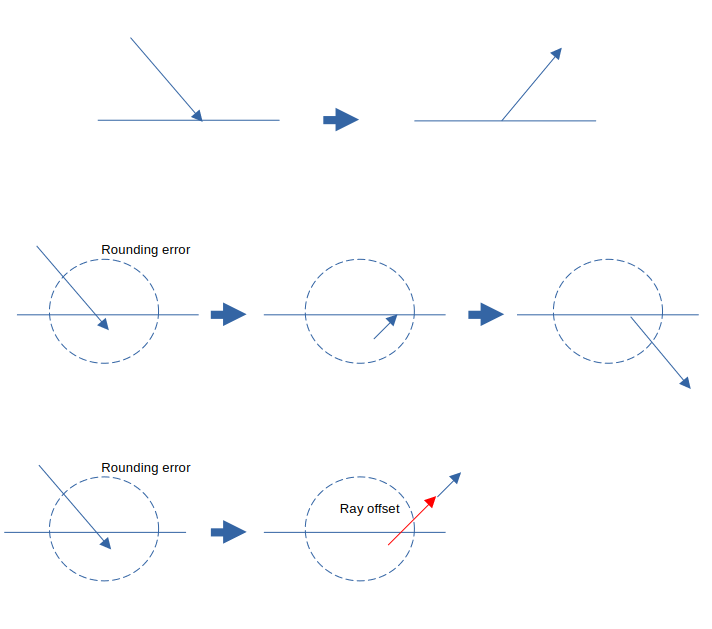}  
    \caption{A typical reflection from a surface: first step ending on the surface and the second step originating from the surface (top). In ray tracing however (middle), rounding errors can cause the second step to start outside its originating volume and thus being scattered from the outer side of the surface, effectively teleporting through a surface. In OptiX, a ray offset is applied to each ray which pushes such escapees (and all rays) back into the volume (bottom).}
    \label{fig:teleport}
\end{figure}

In order to prevent photons from this leakage (or teleport), a small offset is added for each step call along its current direction. As shown in Fig.~\ref{fig:teleport}, even though the origin of the second step is incorrectly in the other volume, the offset pushes it back.

Since EIC-Opticks utilizes the Monte Carlo method to include bulk physics, setting a proper ray offset becomes vital due to the following: in each step the OptiX ray tracing function is called, which returns the surface that photon intersects. Afterwards the Monte Carlo method is utilized to decide which physical process happens to the photon. For example, it can bulk scatter instead of being propagated to the next surface. Yan and his colleagues reported the same phenomenon \cite{yan-2025-spie}.

\begin{figure}[h!]
    \centering
    \includegraphics[width=0.35\textwidth]{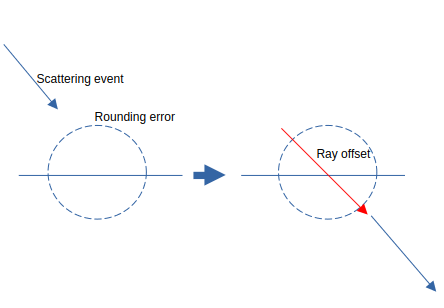}  
    \caption{A ray that is being bulk scattered close to the surface (left) is pushed out from the volume by the ray offset. If the direction of the ray had changed, as in a typical surface reflection, this would not be an issue, since the origin of the next step is moved along the direction of the ray.}
    \label{fig:scatteri}
\end{figure}

In the latter specific case, if the scattering occurs very close to a boundary, the ray offset can push the origin of the next step outside from the volume, as shown in Fig. \ref{fig:scatteri}. Thus, there is an optimal value for the ray offset, if it is too small the photon can leak due to rounding error, while if it is too large it can teleport due to bulk scattering.

\begin{figure}[h!]
    \centering
    \includegraphics[width=0.5\textwidth]{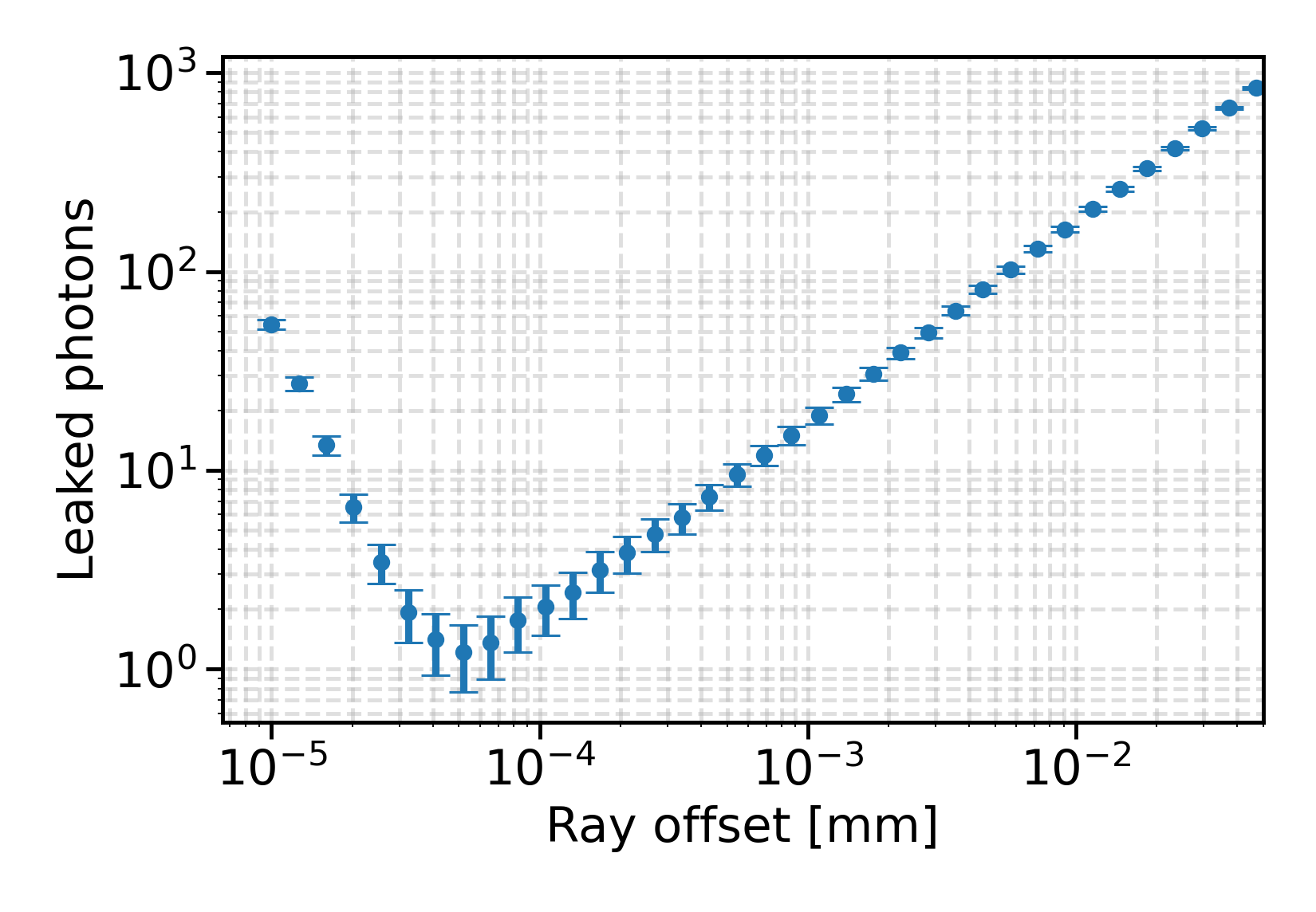}  
    \caption{Fraction of optical photons leaking through a perfectly reflective spherical surface for different ray offsets. The reason for the increase at low offsets is rounding errors, while for large offsets it is caused by bulk scattered events. The optimal ray offset depends on the geometry.}
    \label{fig:leakedphotons}
\end{figure}

In order to quantify photon leakage in the case of EIC-Opticks, we created a simple geometry, consisting of two concentric spheres with radius 15~mm and 25~mm, respectively.  The inner sphere is perfectly reflective on the inside, the outer one detects photons that escape from the inner one. We simulated $10^8$ photons originating from the shared center. The maximum number of reflections were set to 32; the scattering length was 1~m. When setting the ray offset to the default value of 50~$\mu$m, we observed $\sim30$ leaks per $10^6$ photons (30\,ppm). In Fig. \ref{fig:leakedphotons}, the fraction of escaped photons when varying the ray offset setting is shown. One can read off an optimal setting for this specific geometry around 50~nm, for which leakage drops by a factor of 2000 to ${\sim}0.015$\,ppm. Since leakage is geometry dependent, the optimal ray offset will vary. EIC-Opticks provides automated scan utilities so that users can determine and validate the best settings for their specific detector geometry.

\section{Summary}

We introduced EIC-Opticks, a GPU backend for optical-photon transport simulation that extends the Opticks framework to low- and moderate-yield detectors. That was achieved by aggregating photons from thousands of Geant4 events into a single GPU launch. Additionally, we introduced guardrails (e.g. CSG depth checks, scene-epsilon scans) in order to automatize the choice of geometry specific settings. With these tools, users can safely import complex geometries and minimize photon leakage.

We validated EIC-Opticks by simulating electrons entering the ePIC pfRICH detector using Geant4. Comparisons for optical photon transport between EIC-Opticks offloading and standard Geant4 CPU workflow were performed. Agreement in output was confirmed, and a speedup of 161 $\pm$ 3 over single-threaded Geant4 and 10.0 $\pm$ 0.3 over a 20-thread Geant4 configuration was shown.

The event aggregation first implemented in EIC-Opticks reduced the average transport time per photon from \(\sim 60~\mu\mathrm{s}\) (no batching) to \(\sim 20~\mathrm{ns}\) simulating 10,000 events in one batch, a \(\sim 10^3\) throughput gain.

Furthermore, we showed how important is to set the maximum reflections in EIC-Opticks, depending on the specific detector geometry in question. In the case of pfRICH, we showed that a setting of $\sim$24 is optimal. It yields the same number of hits as the Geant4 validation simulation while minimizing the GPU runtime as much as possible. EIC-Opticks provides the tools to carry out this optimization for arbitrary geometries.

We quantified the limitations of EIC-Opticks with respect to tracking error and ``photon leakage''. The latter is a common issue of ray tracing in the NVIDIA OptiX framework. We concluded that the tracking error is far lower than the typical uncertainty in the underlying modeling (e.g., non-perfect mirrors). We additionally showed how photon leakage can be minimized, and we provide scripts in EIC-Opticks so that users can optimize ray tracing's settings for their specific geometry.

By making GPU optical transport efficient in the low-yield regime, EIC-Opticks enables fast, large-scale optical studies for EIC detectors, accelerates geometry/surface scans, and unlocks the production of much larger training sets for machine-learning workflows. Although demonstrated for pfRICH, the approach generalizes to other Cherenkov and scintillator systems --- including medical-imaging simulations ---where per-event photon counts are modest. EIC-Opticks can be installed via Spack or used directly from provided container images.

\section{Acknowledgement}

We gratefully acknowledge Simon Blyth, the developer of Opticks, for his valuable help with this work and for many insightful discussions.  We furthermore thank the pfRICH collaboration for providing us the geometry of the pfRICH detector. This work was supported by the Laboratory Directed Research and Development (LDRD) program under project 26794 of the Brookhaven National Laboratory.
%


\begin{thebibliography}{99}



\bibitem{geant4}
S.~Agostinelli \emph{et al.}, ``GEANT4: A simulation toolkit,'' \textit{Nucl. Instrum. Meth. A} \textbf{506} (2003) 250--303. doi:10.1016/S0168-9002(03)01368-8.

\bibitem{geant4-2016}
J.~Allison \emph{et al.}, ``Recent developments in Geant4,'' \textit{Nucl. Instrum. Meth. A} \textbf{835} (2016) 186--225. doi:10.1016/j.nima.2016.06.125.

\bibitem{opticks-chep2018}
S.~C.~Blyth, ``Opticks: GPU optical photon simulation for particle physics using NVIDIA OptiX,'' in \textit{CHEP 2018} (2018), preprint. Available at: \url{https://simoncblyth.github.io/env/report/opticks-blyth-chep2018-v1.pdf}.

\bibitem{opticks-chep2023}
S.~C.~Blyth, ``Opticks: GPU optical photon simulation via NVIDIA OptiX (7+),'' in \textit{CHEP 2023} (2023), preprint. Available at: \url{https://indico.jlab.org/event/459/papers/11811/files/1068-opticks-blyth-chep2023-v0.pdf}.

\bibitem{optix-2010}
S.~G.~Parker \emph{et al.}, ``OptiX: A general purpose ray tracing engine,'' \textit{ACM Trans. Graph.} \textbf{29}(4) (2010) 66:1--66:13. doi:10.1145/1778765.1778803. Available at: \url{https://research.nvidia.com/sites/default/files/pubs/2010-08_OptiX-A-General/Parker10Optix.pdf}.

\bibitem{eicyellow-2022}
R.~Abdul Khalek, A.~Accardi, J.~Adam \emph{et al.}, ``Science requirements and detector concepts for the Electron-Ion Collider: EIC Yellow Report,'' \textit{Nucl. Phys. A} \textbf{1026} (2022) 122447. doi:10.1016/j.nuclphysa.2022.122447.

\bibitem{spack-sc15}
T.~Gamblin, M.~Legendre, M.~R.~Collette, G.~L.~Lee, A.~Moody, B.~R.~de~Supinski, S.~Futral, ``The Spack package manager: bringing order to HPC software chaos,'' in \textit{SC'15} (2015). doi:10.1145/2807591.2807623. Available at: \url{https://tgamblin.github.io/pubs/spack-sc15.pdf}.

\bibitem{gdml-2006}
R.~Chytracek, J.~McCormick, W.~Pokorski, G.~Santin, ``Geometry Description Markup Language for physics simulation and analysis applications,'' \textit{IEEE Trans. Nucl. Sci.} \textbf{53}(5) (2006) 2892--2896. doi:10.1109/TNS.2006.881062. Available at: \url{https://cds.cern.ch/record/1023367/files/cer-002682496.pdf}.

\bibitem{pbrt-managing-error}
M.~Pharr, W.~Jakob, G.~Humphreys, ``Managing rounding error (PBRT 3e, \S3.9),'' (2018). Available at: \url{https://www.pbr-book.org/3ed-2018/Shapes/Managing_Rounding_Error}.

\bibitem{nvidia-dxr-self-intersection}
N.~Christensen \emph{et al.}, ``Solving self-intersection artifacts in DirectX Raytracing,'' (2023). Available at: \url{https://developer.nvidia.com/blog/solving-self-intersection-artifacts-in-directx-raytracing/}. (Concept applies to OptiX via origin offset / ray\_tmin.)

\bibitem{chatterjee_epic_pid_2025}
C.~Chatterjee, ``Particle identification with the ePIC detector at the EIC,'' \textit{Proc. Sci.} \textbf{469} (2025) 266. doi:10.22323/1.469.0266. (DIS2024 WG6; pre-published 27 Dec 2024.)

\bibitem{yan-2025-spie}
S.~Yan and Q.~Fang, ``Accelerating mesh-based Monte Carlo photon transport simulation using ray-tracing hardware,'' \textit{Proc. SPIE} \textbf{13308} (2025) PC1330806. doi:10.1117/12.3057023.

\bibitem{eic_yellow_report}
R.~Abdul Khalek \emph{et al.}, ``Science requirements and detector concepts for the Electron-Ion Collider: EIC Yellow Report,'' \textit{Nucl. Phys. A} \textbf{1026} (2022) 122447. doi:10.1016/j.nuclphysa.2022.122447.

\bibitem{docker-merkel2014}
D.~Merkel, ``Docker: Lightweight Linux containers for consistent development and deployment,'' \textit{Linux Journal} \textbf{239} (2014) 2. Available at: \url{https://www.linuxjournal.com/content/docker-lightweight-linux-containers-consistent-development-and-deployment}.

\bibitem{pbrt3e}
M.~Pharr, W.~Jakob, G.~Humphreys, \textit{Physically Based Rendering: From Theory to Implementation}, 3rd ed. (Morgan Kaufmann, San Francisco, 2016). ISBN 978-0-12-800645-0.

\bibitem{epic-pid-2024}
Y.~Song \emph{et al.} (ePIC Collaboration), ``Particle identification with the ePIC detector at the EIC,'' \textit{arXiv}:2410.20410 (2024). Available at: \url{https://arxiv.org/abs/2410.20410}.

\bibitem{pfrich-cdr-2024}
ePIC pfRICH Subsystem Collaboration, ``A Proximity-Focusing RICH for the ePIC Experiment -- Conceptual Design Report (v1.1),'' Tech. Rep. (2024). Available at: \url{https://indico.bnl.gov/event/19827/attachments/48201/81847/EPIC-pfRICH-CDR.v1.1.pdf}.
\end{thebibliography}
\end{document}